\documentclass[9pt,twocolumn,twoside]{osajnl}

\journal{ao} 

\setboolean{shortarticle}{false}

\title{Building hybridized 28-baseline pupil-remapping photonic interferometers for future high resolution imaging}

\author[1,*]{Nick Cvetojevic}
\author[2,3,4]{Barnaby R. M. Norris}
\author[5]{Simon Gross}
\author[6]{Nemanja Jovanovic}
\author[5]{Alexander Arriola}
\author[7]{Sylvestre Lacour}
\author[8,9,10]{Takayuki  Kotani}
\author[11]{Jon S. Lawrence}
\author[5]{Michael J. Withford}
\author[2]{Peter Tuthill}

\affil[1]{Université Côte d'Azur, Observatoire de la Côte d'Azur, CNRS, Laboratoire Lagrange, France}
\affil[2]{Sydney Institute for Astronomy, School of Physics, Physics Road, University of Sydney, NSW, Australia}
\affil[3]{AAO-USyd, School of Physics, University of Sydney, NSW 2006, Australia}
\affil[4]{Sydney Astrophotonic Instrumentation Laboratories, Physics Road, University of Sydney, NSW, Australia}
\affil[5]{MQ Photonics Research Centre, Department of Physics and Astronomy, Macquarie University, NSW 2109, Australia}
\affil[6]{Department of Astronomy, California Institute of Technology, 1200 E. California Blvd, Pasadena, CA, 91125, USA}
\affil[7]{Observatoire de Paris, LESIA, 5 place Jules Janssen, 92190 Meudon, France}
\affil[8]{Astrobiology Center, NINS, 2-21-1 Osawa, Mitaka, Tokyo 181-8588, Japan}
\affil[9]{National Astronomical Observatory of Japan, NINS, 2-21-1 Osawa, Mitaka, Tokyo 181-8588, Japan}
\affil[10]{Department of Astronomy, School of Science, The Graduate University for Advanced Studies (SOKENDAI), 2-21-1 Osawa, Mitaka, Tokyo, Japan}
\affil[11]{Australian Astronomical Observatory, Faculty of Science and Engineering, Macquarie University, NSW 2109, Australia}

\affil[*]{Corresponding author: nick.cvetojevic@oca.eu}

\doi{\url{https://doi.org/10.1364/AO.422729}}

\begin{abstract}
One key advantage of single-mode photonic technologies for interferometric use is their ability to easily scale to an ever increasing number of inputs without a major increase in the overall device size, compared to traditional bulk optics. This is particularly important for the upcoming ELT generation of telescopes currently under construction. We demonstrate the fabrication and characterization of a novel hybridized photonic interferometer, with 8 simultaneous inputs, forming 28 baselines, the largest amount to-date. 
Utilizing different photonic fabrication technologies, we combine a 3D pupil remapper with a planar 8-port ABCD pairwise beam combiner, along with the injection optics necessary for telescope use, into a single integrated monolithic device. We successfully realized a combined device called Dragonfly, which demonstrates a raw instrumental closure-phase stability down to $0.9^{\circ}$ over $8\pi$ phase piston error, relating to a detection contrast of $\sim6.5\times 10^{-4}$ on an Adaptive-Optics corrected 8-m telescope. This prototype successfully demonstrates advanced hybridization and packaging techniques necessary for on-sky use for high-contrast detection at small inner working angles, ideally complementing what can currently be achieved using coronagraphs. 
\end{abstract}

\setboolean{displaycopyright}{true}

\begin{document}

\maketitle

\section{Introduction}
\label{sec:intro}

Direct imaging of exoplanets, where a star and nearby planet are separately resolved at an image plane, promises to provide critical answers to questions of planetary formation and evolution. However, most of the exoplanets imaged thus far have been limited to wide separations \cite{B12}. This is because of the challenging nature of making high contrast observations at very small spatial scales due to the glare of the host star, and the limited angular resolution of large telescopes. Although coronagraphs offer the highest possible contrast, due to residual wavefront aberrations after Adaptive Optics (AO) correction, it is difficult to achieve this close to the star ($1-3\lambda / D$) \cite{guyonreview}. It is this inner $3\lambda / D$ region that is of critical importance to understanding planetary formation as it corresponds to  solar-system scales, where interactions between dust populations and proto-planetary bodies occur \cite{B3}. Further, this region overlaps with existing transit and radial-velocity planet detections, and is statistically more likely to contain possible planets \cite{snowline}. To address this parameter space, interferometric techniques such as aperture-masking are used, wherein the pupil of a large telescope is divided into a number of small sub-pupils using an opaque mask placed at the pupil plane. This essentially turns the telescope into a sparse interferometer array \cite{CP_OGpaper,B14}. From the resulting interference pattern, it is possible to extract observables, such as the visibility and the closure phase, with the latter being highly robust to residual phase aberrations. By exploiting closure phases, the telescope's diffraction-limited performance can be recovered\cite{CP}. 

However, this approach has its own limitations. Most of these masks have the sub-aperture positions be non-redundant to reduce noise from repeating baselines, which severely limits the number of sub-apertures on the mask. This results in low throughputs, as only a fraction of the total pupil is transmitted by the mask. For example, a commonly used 9-hole mask has a throughput of only $\sim 12 \%$, restricting the use of the technique to bright targets only. A further limitation is imposed by the non-zero size of the sub-apertures, which allows for phase variations that affect the measured observables by decreasing the signal-to-noise of the instrument. 

Photonic technologies offer a unique ability to overcome most of these limitations, as part of a concept known as photonic pupil-remapping (PRM) interferometry \cite{DflyNem,FIRST1}. Instead of using an aperture mask, the pupil of the telescope can be sampled by a single-mode optical waveguide, which then coherently remaps the 2D pupil into a linear array, forming the basis of a 1D interferometer. This is achieved by routing the light in 3D, either through the use of waveguides, or optical fibres. This has a number of distinct advantages. Since the light guides are single-moded, any phase-variation across a single sub-aperture are inherently filtered out, effectively removing one of the the signal-to-noise limitations found in aperture masking \cite{FLOUR1, SM_benefit}. Furthermore, because the output array is 1D, it is suitable to interface with photonic chip-based beam-combiners \cite{Gravity_photonics, Gravity_photonics2, solGel}, which have been successfully used in large baseline interferometers such as PIONEER \cite{pionier1} and GRAVITY \cite{Gravity1}. These devices perform another major task that is difficult to replicate in bulk optics. By using a series of cascaded 50/50 waveguide couplers, the beam-combiner transforms what is essentially an imaging problem (fringes from different baselines imaged by a detector) into a photometric one. This type of beam-combining architecture reconstructs the fringe by using four discrete waveguide outputs per baseline. Further, as the output is a 1D array of waveguides, it is possible to spectrally disperse using bulk optics in the perpendicular direction, and obtain a broad wavelength coverage, or alternatively, keeping the light on-chip by dispersing using integrated photonic spectrographs \cite{AWG1,AWG2,AWGNem,AWGPradip, GlenAWG1, GlenAWG2}. 

These benefits combine to create a platform ideally suited for high-angular resolution science. The interferometric self-calibrating  nature of closure phase suppresses the impact of AO residuals, allowing for detections of companions well below the formal diffraction limit of the telescope \cite{FIRST1}. Furthermore, because the detection contrast limit is generally a function of the instrumental closure phase stability (detailed in Section \ref{sec:discussion}), the stability offered by an integrated photonic platform can achieve high-contrast detections at $10^{-3}-10^{-4}$.

We present a novel hybridized device which combines both the 3D pupil remapping photonic chip (PRM chip), and the planar beam-combination chip, into a single monolithic device called Dragonfly. Further, we describe how the chips are packaged and bonded, along with a pre-aligned microlens array for segmenting and injecting a telescope pupil, into a robust and stable assembly. We present the laboratory characterization of the device, including the throughput and the closure phase stability, a key metric in assessing on-sky performance. 


\section{Photonic components}
\label{sec:photonics}

There were two custom photonic circuits used in the overall device. The first is a 3D PRM chip as used in \cite{DflyNem, DflyBarn, DflyAlex}, fabricated using Ultra-fast Laser Inscription (ULI), which samples the telescope pupil using 8 waveguides in two dimensions, and remaps it into a linear array of evenly spaced waveguides at the output. Critically, the waveguide routing is designed to ensure there is no additional differential optical-path delay (OPD) between the inputs, thus preserving near-perfect coherence of the resulting interferometric baselines \cite{DflyNed}. Since all the waveguides are in a monolithic glass block, environmentally induced OPD errors are greatly reduced when compared with alternative remapping techniques, such as optical fibers. This is because environmental effects, such as temperature or strain, are felt more globally by all waveguides, instead of individually by fibres which can lead to OPD drift in any given baseline. 

The second is a pairwise beam-combination chip fabricated using Silica-on-Silicon photo-lithography. This chip takes 8 inputs and interferes them forming 28 distinct baselines, encoding the resulting fringes from each baseline in 4 outputs. This architecture (commonly referred to as an ABCD pairwise combiner) is identical in form to the photonic chips used in the GRAVITY \cite{Gravity_photonics} and PIONIER\cite{pionier1} instruments at the VLTI, albeit with double the number of inputs.

\subsection{3D Pupil-remapping chip}
\label{sec:remapper}

The PRM chip consists of 8 separate single-mode waveguides optimized for the astronomical H-band ($\sim 1.6~\mu m$). The single-mode waveguides were inscribed inside a monolithic block of boro-alumino-silicate glass (Corning Eagle2000) using a tightly-focused femtosecond-pulsed laser, which creates a positive refractive index change inside the glass through a nonlinear photon absorption process \cite{ULI1,ULI2,ULI3,ULI4}. The waveguides were written with a $800~nm$ wavelength, $5.1~MHz$ repetition rate Ti:sapphire laser, at a pulse energy of $90 nJ$. The inscription laser is focused $350~\mu m$ below the top surface, to a sub-micron spot-size using a $100\times$~$1.25~NA$ oil immersion microscope objective (Zeiss N-Achroplan). After writing, the waveguides were thermally annealed to smooth the refractive index profile by removing the unwanted outer cladding, and remove internal stresses in the glass to yield waveguides with a single-mode cut-off of $1050~nm$ \cite{DflyAlex}.  

\begin{figure}[ht!]
\centering\includegraphics[width=0.7\linewidth]{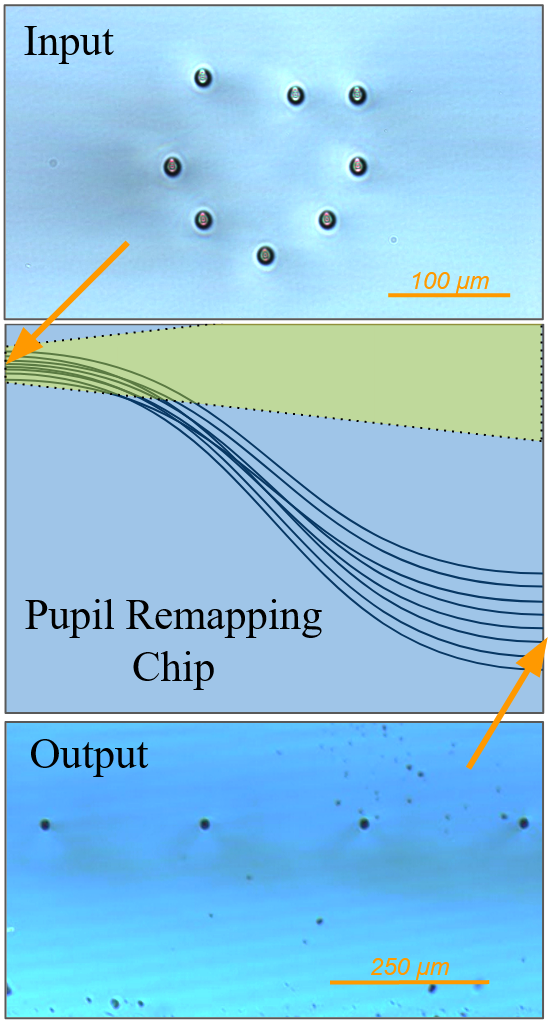}
\caption{The pupil remapping chip routes 8 single-mode waveguides from a 2D pupil plane (top) to a $250~\mu m$ spaced linear array at the output (bottom). The outputs are offset laterally by $5~mm$ in order to `side-step' stray light (yellow cone) in the bulk from imperfect coupling at the input (center).}
\label{fig:remap}
\end{figure}

The glass block was then translated with respect to the laser beam using computer-controlled precision air-bearing stages at a velocity of $500~mm/min$ allowing the laser to sculpt the desired waveguide circuitry in three dimensions. The 2D waveguide arrangement at the input was chosen to sample a typical telescope pupil (avoiding the shadow of the secondary mirror) in a non-redundant manner, as shown in Figure~\ref{fig:remap} (top). The waveguide separation was matched to our choice of commercially available microlens array ($30~\mu m$ pitch). The waveguides were remapped to a linear array at the output, with a $250~\mu m$ pitch (Fig.~\ref{fig:remap} bottom).  

Apart from the 2D-to-1D remapping, the chip circuitry includes a large `side-step’ formed by a cosine S-bend, where the waveguide position is translated laterally by $5~mm$ while maintaining matched path-lengths. This is a critical feature that ensures any light not perfectly coupled into the input waveguides does not cause unwanted interference at the output, which has been shown to negatively impact measurement accuracy \cite{DflyBarn}. The `side-step’ results in a minimum waveguide radius of curvature of $29~mm$, minimizing bend-losses. The final single-mode waveguides had a mode-field (MFD) of $10 \times 8.5~\mu m $ in the two axes. The entire photonic chip measures $30~mm$ in length by approximately $10~mm$ in width, with a thickness of $1.1~mm$.

\subsection{Beam-combiner chip}
\label{sec:IOchip}

\begin{figure}[ht!]
\centering\includegraphics[width=\linewidth]{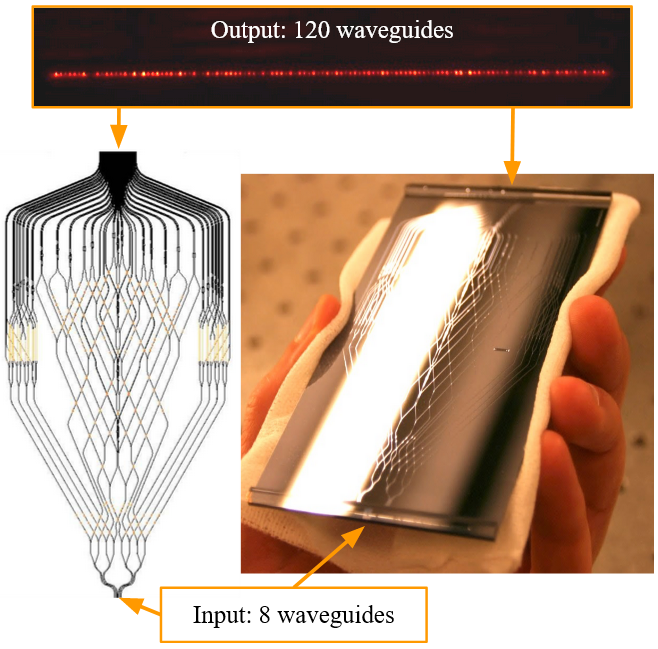}
\caption{A schematic and image of the 8-input ABCD beam combining chip. The inputs are interfered entirely on-chip, with the resulting 120 output waveguides routed to the chip end face (top). }
\label{fig:IOchip}
\end{figure}

To obtain interferometric measurements of the remapped telescope pupil, the light is passed to a dedicated planar beam-combination chip, shown in Figure~\ref{fig:IOchip}. This chip was fabricated using a UV photo-lithographic process where a doped-silica waveguide core of higher refractive index is surrounded by a cladding of lower refractive index pure silica. This is achieved by exposing a UV-photosensitive core layer through a mask that contains the desired waveguide circuitry for the whole chip, then subsequently chemically etching the layer to produce the waveguide cores, before capping the chip with an upper cladding. Due to the manufacturing process, the entire chip is planar (all the waveguides sit in the same vertical plane, unlike the previously described PRM chip).

The chip has 8 evenly spaced input waveguides at $250~\mu m$ pitch. Each of the inputs are initially split using a Y-junction to pick off $\sim 5\%$ of the total light into photometric channels, allowing for the instantaneous measurement of coupling efficiency. Next, each of the input waveguides are divided up further into 7 equal channels and interfered with each other in a pairwise fashion, creating 28 unique interferometric baselines. Each baseline combination (of two input waveguides) is further split in two and interfered using two directional couplers, with one having a $\pm \pi /2$ phase offset. Thus, each baseline has 4 interferometric output waveguides (an ABCD combiner \cite{ABCD}), making 112 output waveguides for all 28 baselines. In total, the chip has 120 output waveguides, spaced equally by $80~\mu m$.


\section{Hybridization and assembly}
\label{sec:assembly}

To create a practical and robust photonic interferometer from the components listed in Section~\ref{sec:photonics}, the beam-combiner and PRM were bonded and packaged with a microlens array to focus the light from the pupil into the device. With each component having 6-degrees of freedom for alignment (X-Y-Z translation + Pitch, Yaw, Roll), and with a substantial accuracy requirement to avoid additional losses, independently aligning each component was deemed unfeasible in any practical on-sky scenario. Thus, each component was carefully aligned, bonded, and packaged to form a monolithic device.

\begin{figure}[ht!]
\centering\includegraphics[width=\linewidth]{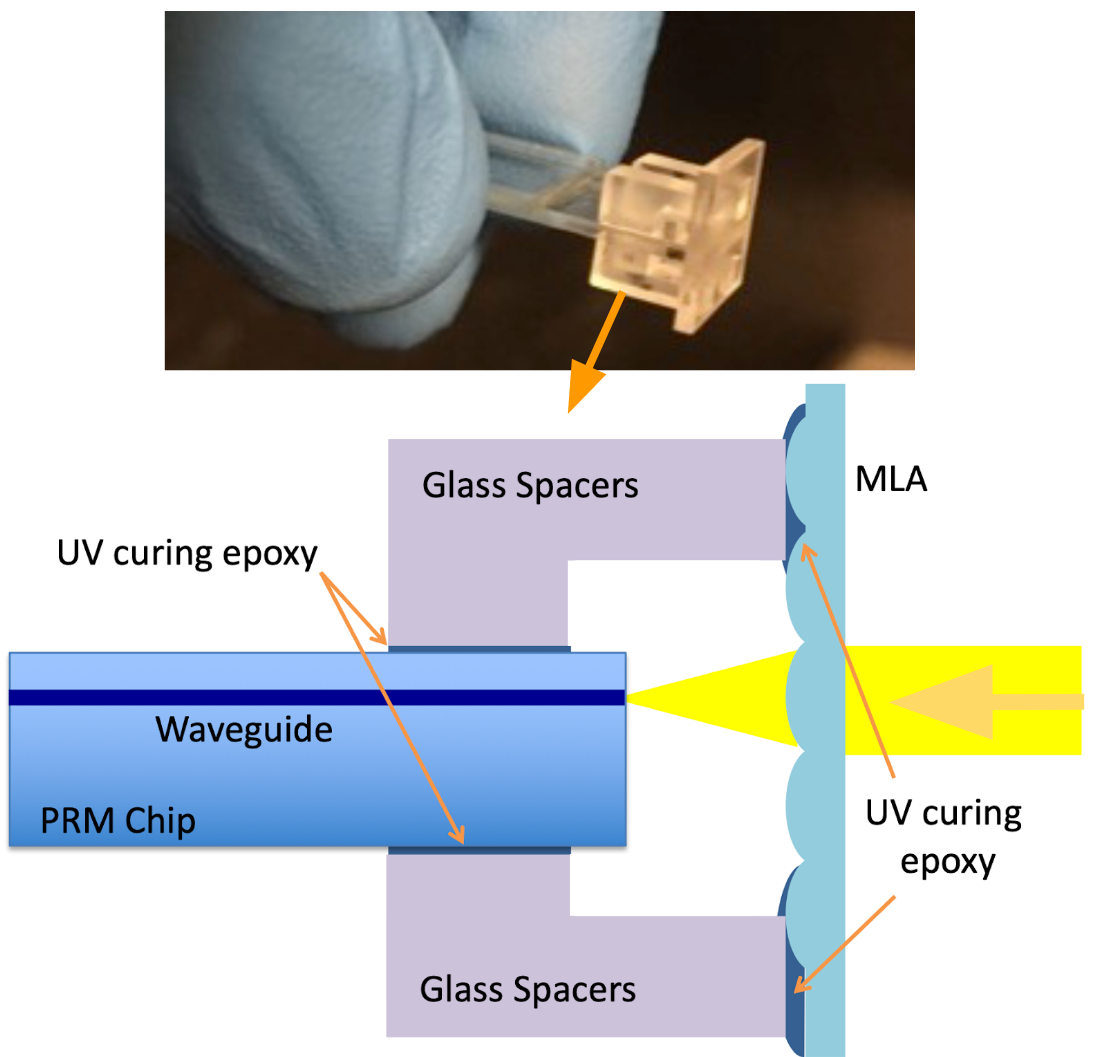}
\caption{A schematic demonstrating the components for bonding the MLA to the PRM chip. The incoming collimated light from the right-hand side hits each lenslet, which focuses it into a waveguide in the chip. Once the bonding procedure is complete, the components are very rigid and can be handled easily (top).}
\label{fig:MLAbonding}
\end{figure}

The first and most difficult step was the alignment of the microlens array (MLA) used for injection with the PRM chip. The MLA features $30~\mu m$ diameter lenslets on a hexagonal grid with a $30~\mu m$ pitch, a focal length of $96~\mu m$, and a numerical aperture (NA) of 0.16. The fused silica substrate of the MLA measured $10 \times 10$~mm in area with a $1$~mm thickness. Because of the short focal length of the lenslets, the MLA was oriented ‘backwards,’ with the flat substrate facing the incoming collimated beam, and the convex side facing the photonic chip. Unfortunately, simply placing UV curing epoxy between the MLA and photonics would not work, as this changes the refractive index of the glass-air interface required for the MLAs to function at specification. Thus, a more complicated bonding method was used where two glass ’L’ shaped spacers were attached to the top and bottom of the photonic PRM and the front of the MLA chip (see Figure \ref{fig:MLAbonding}).

The alignment procedure first required the angle of the MLA and PRM end-faces to be made parallel. To achieve this, the angles of both were adjusted while the Fresnel back-reflected reference beams were co-aligned in collimated space. This was critical, as any angular misalignment between the MLA and reference beams will result in a phase ramp across the input when the system is aligned for optimum injection efficiency. Once the MLA was perfectly parallel to the chip, the remapping chip was brought
into optimal alignment by being back-illuminated via 8x butt-coupled optical fibers (in a V-groove array) at the output face. The distance between the PRM and MLA was adjusted by measuring the gap using a calibrated vision system. In the next step, the pupil of the MLA was imaged onto a detector, and the light emerging from all the waveguides was centred with respect to the pupils of the lenslets by translating the chip laterally with respect to the MLA using a high-precision, piezo actuated 6-axis translation stage. Once the initial alignment was optimized, the L-spacers were placed on top of the chip and attached with UV curing epoxy (Norland NOA61) on the mating faces. 

\begin{figure}[ht!]
\centering\includegraphics[width=0.95\linewidth]{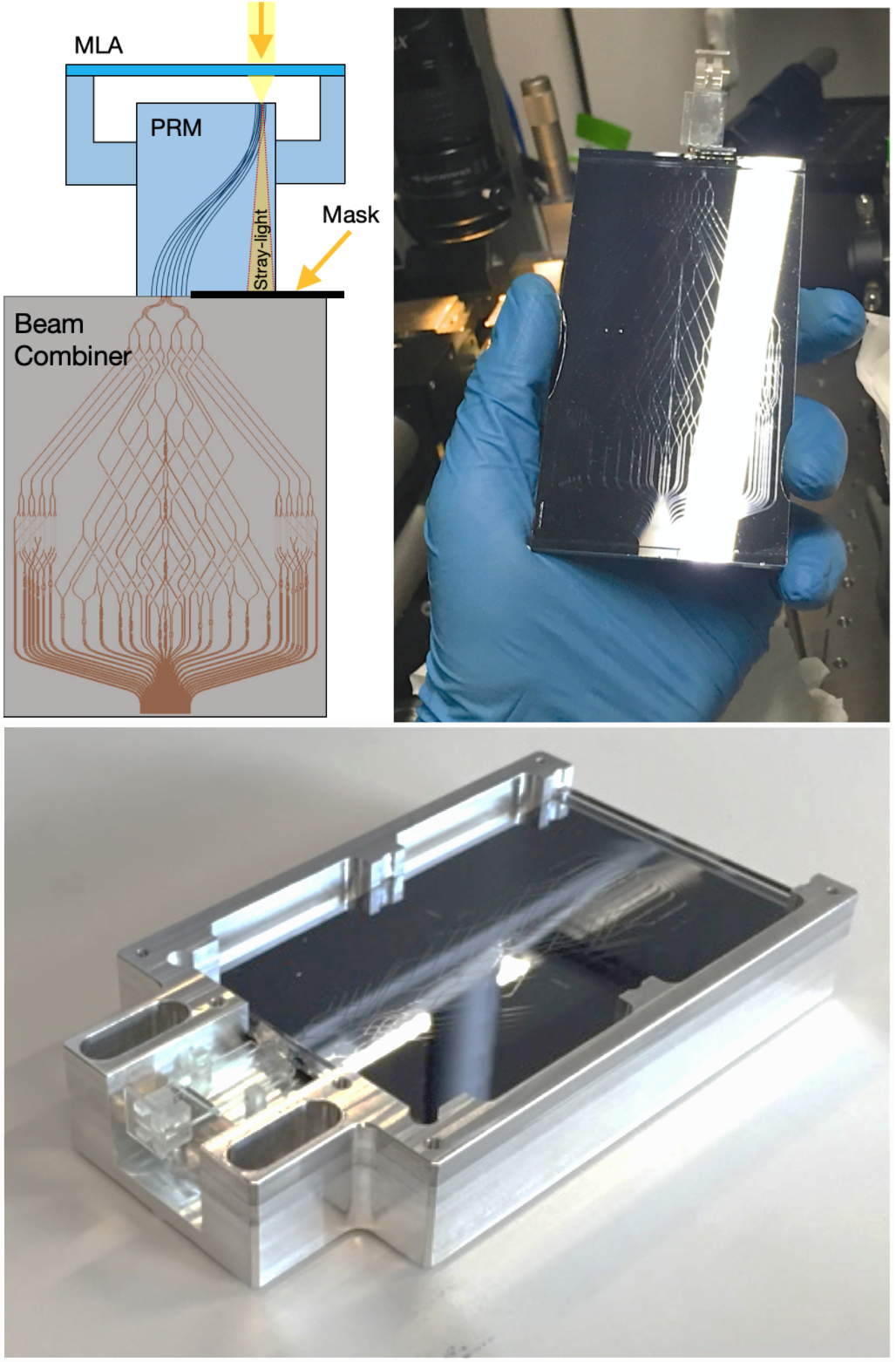}
\caption{A schematic (not to scale) of the fully bonded photonics (top-left). The PRM-MLA assembly shown in Fig. \ref{fig:MLAbonding} is aligned and bonded to the beam-combining chip. A thin mask is placed across half the PRM output to block any stray light from entering the subsequent chip. After bonding, the chip assembly is rigid and can be handled easily (top-right), and is then placed in a custom housing for easy on-telescope deployment (bottom).}
\label{fig:fullbonding}
\end{figure}

With the MLA-PRM chip bonded, the beam-combiner chip was also aligned and bonded to the upstream assembly (shown in Figure~\ref{fig:fullbonding}). The MLA-PRM assembly was first aligned to a collimated input beam from a broadband NIR source and locked in place. The beam-combiner chip was placed on a high-precision, piezo actuated 6-axis translation stage, moved into position downstream of the MLA/PRM assembly, and adjusted to maximize coupling. The coupling efficiency was monitored using a NIR InGaAs detector (Xenics Xeva-1.7-640) imaging the beam-combiner output face. Prior to bonding, a few-micron thick brass mask was placed in between the two chips to block any stray light from the PRM chip entering the beam-combiner, offset laterally so as not to block any of the waveguides. No further machining of the chips was done to accommodate the mask, as the impact of the thin foil did not cause any measurable increase in coupling loss between the chips. 

Once the entire unit was bonded together, it was placed in a custom housing made from Computer Numerical Control (CNC)-machined Aluminum. This housing had support structures and pillars for the assembly to attach to so there were no overhanging segments that could vibrate. Once bonded into the housing, the entire device was rigid and robust so it could be easily positioned at the telescope pupil plane. The housing had attachment points to bolt onto standard translation stages, as well as a perspex cover to protect form dust and damage.


\section{Characterization}
\label{sec:characterisation}

\subsection{Experimental setup}

\begin{figure*}[ht!]
\centering\includegraphics[width=0.79\linewidth]{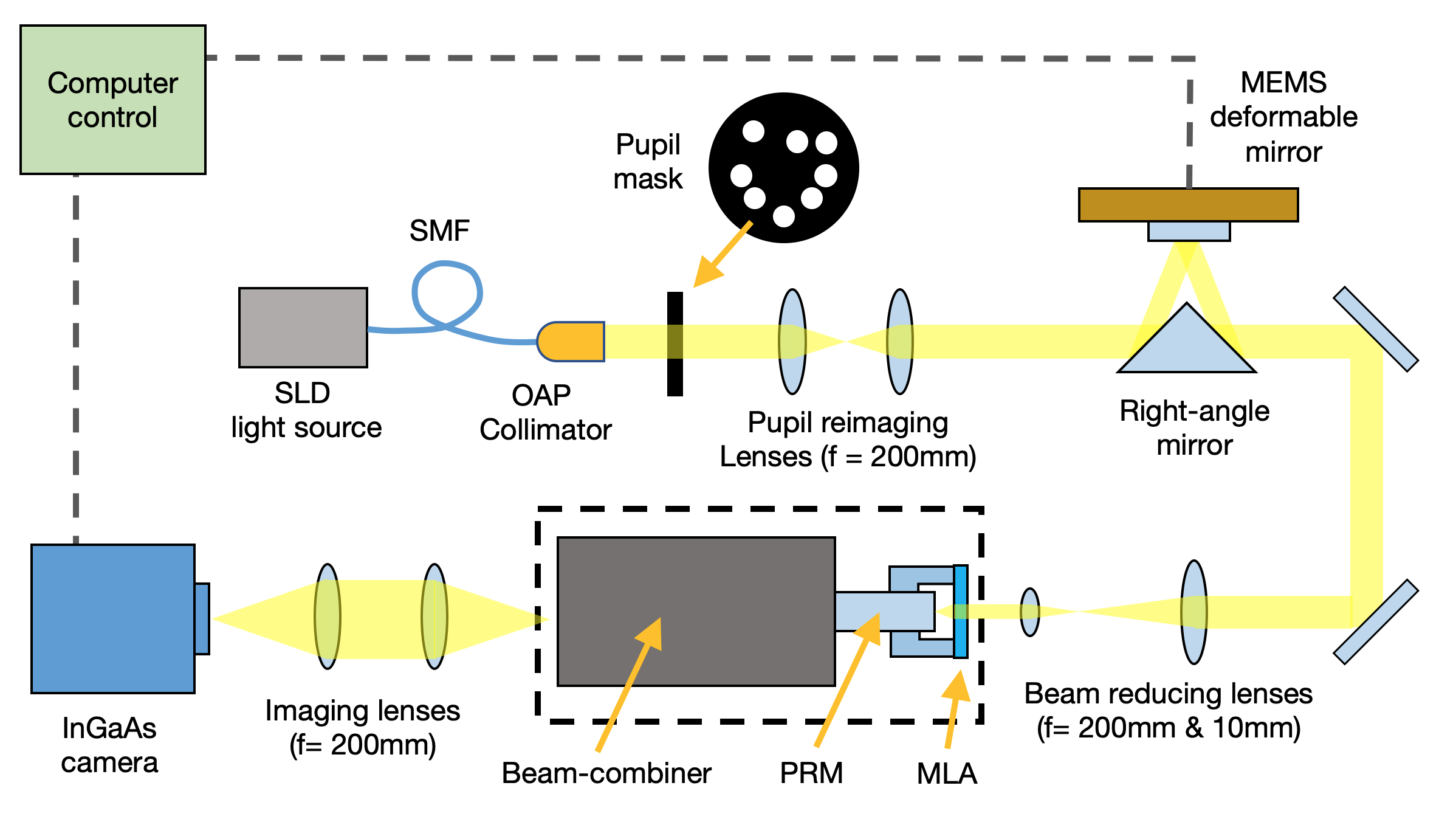}
\caption{Schematic of the characterization test bench. The beam from an SMF-coupled SLD light source was collimated via an off-axis parabolic reflective collimator, and passed through a pupil-mask. The pupil was reimaged onto a segmented MEMS deformable mirror, and again onto the front face of the photonic assembly. The output of the chip was imaged onto an InGaAs camera for data acquisition and the closed-loop injection optimization software.}
\label{fig:setup}
\end{figure*}

The characterization setup was similar to that used in previous work on the PRM chips \cite{DflyBarn}, and is shown in Figure \ref{fig:setup}. A collimated beam from a NIR superluminescent diode (SLD) source was used to form a pupil on a 1-inch brass mask with $8\times~600~\mu m$ holes in an arrangement matching that of the PRM input waveguide layout. The pupil was then re-imaged onto a 37-element segmented MicroElectroMechanical System (MEMS) deformable mirror (IrisAO PT-111), which provided precise tip, tilt, and piston control for each of the pupil sub-apertures. The pupil was again imaged and demagnified by $20\times$ to form the final pupil plane at the entrance of the assembly (the bonded MLA front-face). The full assembly was attached to a precision X-Y-Z translation stage, with tip-tilt, and aligned such that the (now $30~\mu m$ diameter) mask holes overlap precisely with the MLA lenslets to focus into the PRM waveguides. Because imaging through the MLA is not possible with the bonded assembly, it was translated into a rough alignment by back illuminating two of the photometric ports at the chip output and monitoring the reflection of the collimated sub-pupils on the upstream mask. This was enough to get the assembly close to the correct position and focus, and thus could switch to monitoring the beam-combiner output until the flux out of all 8 photometric ports was optimized. The output of the chip was imaged onto a 640x512 pixel, NIR InGaAs detector (Xenics Xeva-1.7-640) using a 1-to-1 imaging system ($f=200~mm$, $2"$ achromatic doublets), with a typical output shown in Figure~\ref{fig:IOchip} (top right).

The MEMS allows for direct independent control of the upstream wavefront of each sub-aperture, in terms of piston, tip, and tilt, over a range of $\pm 8~\mu$m and $\pm 10~$mrad respectively. The MEMS is initially used for optimizing the injection into each input (by tuning tip-tilt), but then can be used to ``\textit{turn-on}'' or ``\textit{turn-off}'' each input by switching between the optimized tip-tilt value and $\pm 5$~mrad (end of the travel range) value so that no light is coupled into the waveguide. Importantly, this is only possible due to the sidestep structure in the PRM, as without it the now uncoupled light will add to unwanted interference error in the bulk. Thus, which inputs are active at any given time can be reconfigured on the fly and iterated in a characterization loop. Further, the piston control allows for the accurate scanning of the fringes for all baselines independently.

\subsection{Throughput}

The throughput for each element of the device is presented in Table \ref{tab:throughputs}. Coupling losses occur at the MLA-PRM interface due the mismatch between MLA focal spot and waveguide mode. The exact value varies depending on how much of the total area of each lenslet is filled by the pupil sub-aperture, defined in our setup by the upstream mask. For the $550~\mu m$ sized mask holes used in our measurements, the calculated coupling efficiency into the waveguides is $77.0\%$.

\begin{table}[htbp]
\centering
\caption{\bf Throughputs for each component in the assembly}
\begin{tabular}{cc}
\hline
Part & Throughput (\%) \\
\hline
MLA injection & $77.0$  \\
PRM chip & $72 \pm 1$  \\
PRM-BC chip coupling & $\sim 93$  \\
BC Chip & $\sim 59$ \\
Fresnel losses & $84$ \\
\hline
Total throughput & $\sim 26$ \\
\hline
\end{tabular}
  \label{tab:throughputs}
\end{table}

The internal transmission of the PRM chip was measured in the same method used for previous devices \cite{DflyAlex, Meany_2014}, and was measured to be $72 \pm 1~\%$ at $1550$~nm providing an upper bound for the waveguide propagation losses of $\sim 0.3~dB/cm$ \cite{Meany_2014}. The internal transmission includes losses due to propagation, bend losses, and absorption caused by impurities of the substrate material ($\sim 20\%$ \cite{Meany_2014}). 

The coupling loss between the PRM and beam combining chip is estimated to be $\sim 7\%$, by calculating the mode-overlap integral between the two dissimilar waveguides. The total throughput map of the beam-combiner chip was difficult to determine accurately, as decoupling the effects of injection losses, transmission, internal interference, and coherent cross-coupling proved challenging, and varied greatly with each output. However, an overall estimate of the averaged beam-combiner chip throughput is $\sim 59~\%$.

Additionally, Fresnel reflection losses totalling 16\% occur at the uncoated surfaces of the MLA and chip input and output face. Thus, an upper bound for the end-to-end transmission efficiency for the entire assembly (assuming perfect injection) is $\sim 26 \%$.

\subsection{Closure Phase Stability}

Data was acquired using the tip/tilt movement of individual MEMS segments to switch desired waveguides ‘\textit{on}’ or ‘\textit{off}’ (i.e. by tilting the sub-beam far enough away from the waveguide that no light is coupled), and using the piston movement to induce a variable phase delay in chosen baselines. To fully characterize the chip, three types of data sets are typically acquired. First are single waveguide scans, where one waveguide of the PRM is illuminated. Next are baseline scans where each of the 28 baselines in turn are enabled by switching on the relevant pair of waveguides. A phase ramp is applied to each waveguide sequentially by adding piston with the MEMS, resulting in a data-cube containing the output images as a function of piston position ($\Delta~OPD$). Finally, an “all-on” scan is done, where all 8 waveguides were switched on, and individual waveguides had phase ramps applied. This best simulates on-sky data collection, where all baselines are measured simultaneously with all inputs suffering from phase error induced by the atmosphere.

\begin{figure*}[ht!]
\centering\includegraphics[width=\linewidth]{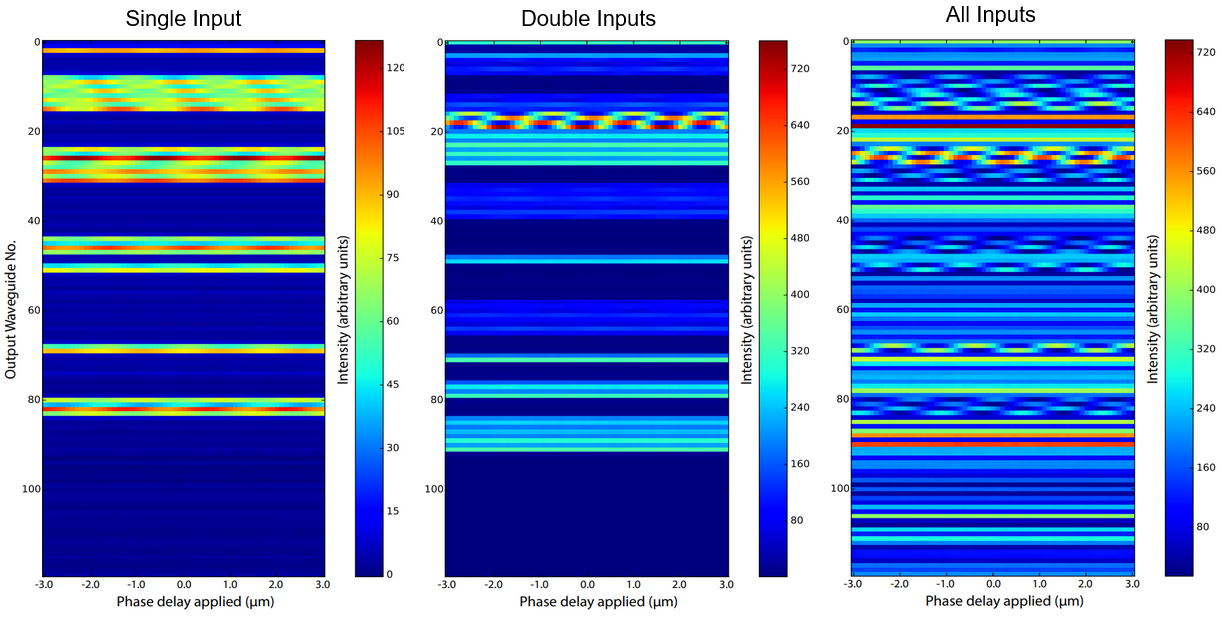}
\caption{The output response of the beam-combiner chip for all outputs (y-axis) as a function of upstream OPD piston (x-axis). For the case when only a single input is illuminated (left) only the outputs that pertain to that baseline have signal in them. When only two waveguides are activated and one input is pistoned (center), the expected fringing is measured on the 4 outputs that correspond to the specific baseline. Lastly, when all the inputs are activated and a single input pistoned, we measure the desired fringes on all 7 baselines that include that input (right).}
\label{fig:waterfall}
\end{figure*}

The encircled flux of each of the output waveguides was measured for every applied step in phase delay. The single waveguide scans were used to measure the transmission coefficients of the waveguides (in terms of intensity), and the baseline scans (where a known phase ramp is added to each baseline in turn) was used to measure the phase relationship between the baseline phase and each of the relevant ABCD outputs. To test the closure phase precision, the all-on data set was used, with the phase of each baseline being determined by fitting a sine function to all of the ABCD outputs (using a non-linear least squares fit Levenberg-Marquardt algorithm). The previously measured intensity-phase relationship and transmission coefficients for each output are used to calibrate out variations between the baselines. The resulting measured phases from closing triangles of baselines were added to produce closure phases. Figure \ref{fig:waterfall} shows the variation in encircled flux for all output beam-combiner waveguides as the phase is pistoned on a single input waveguide. In the figure, three cases are shown; only one input active, one baseline active, and all inputs active. The ABCD output baselines that include said waveguide modulate through the resulting fringe as the input phase is ramped, with the remaining outputs being unaffected.

\begin{figure}[ht!]
\centering\includegraphics[width=0.9\linewidth]{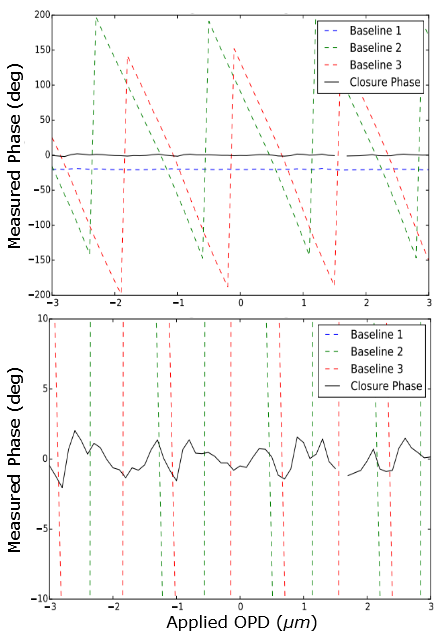}
\caption{The closure phase stability of a typical triangle when one baseline is pistoned over a 6~$\mu m$ OPD piston ($8\pi$). The dotted coloured lines show the individual phases of the three baselines forming the closure triangle, while the solid black line is their sum (the closure phase). The mean closure phase (of $-61.3^{\circ}$) has been subtracted for clarity of plotting. The bottom plot shows a smaller range of phase angles so that the small closure phase variations are more visible.}
\label{fig:bestCP}
\end{figure}

\begin{figure}[ht!]
\centering\includegraphics[width=0.9\linewidth]{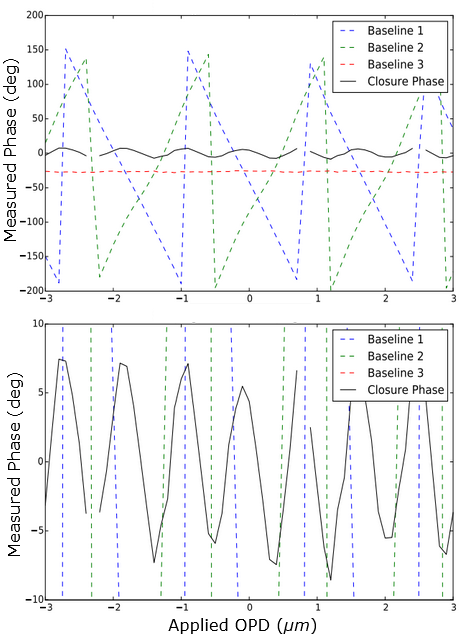}
\caption{The closure phase stability of an unstable triangle when one baseline is pistoned. This closure triangle exhibited poor closure phase stability ($CP\sigma =4.7^{\circ}$). The periodic nature of the instability is primarily due to the effect of cross-talk between the waveguides.}
\label{fig:badCP}
\end{figure}

\begin{figure}[ht!]
\centering\includegraphics[width=0.9\linewidth]{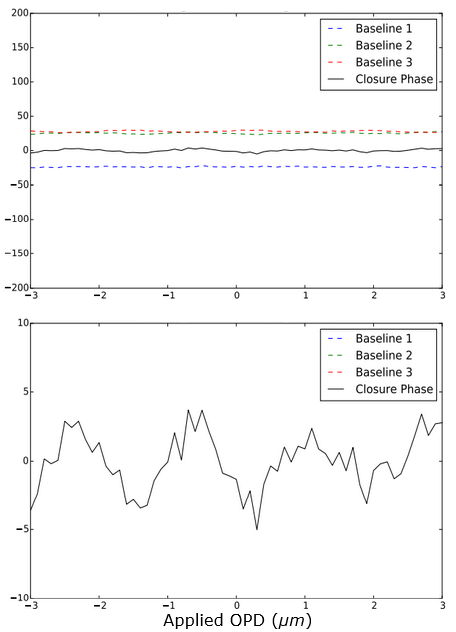}
\caption{The closure phase stability when no baseline is pistoned. This triangle should be exceptionally stable as there are no changes actively made to the input waveguides that make up the closure triangle (typically, $CP\sigma =\sim0.66^{\circ}$), however this particular case showed poor closure phase stability, with $CP\sigma =2.0^{\circ}$.}
\label{fig:badCP_nopiston}
\end{figure}

The measurements of closure phase presented in this section were conducted with all the waveguides in the ‘\textit{on}’ position, but only a single input pistoned, with a typical datacube shown in Figure  \ref{fig:waterfall} (far right).
The closure phase measured from the beam-combiner should remain constant throughout all measurements in the ideal case. It was found that for some of the best baseline triangles, the standard deviation of the closure phases ($CP\sigma$) when one baseline of the triangle is pistoned through 8$\pi$ radians is $CP\sigma = 0.9^{\circ}$, with the measured closure phase shown in Figure \ref{fig:bestCP}. However, some triangles suffered from much larger closure phase instabilities ($CP\sigma = 4.7^{\circ}$), shown in Figure \ref{fig:badCP}. This variation in closure phase stability was observed both in triangles where a constituent baseline was pistoned, as well as in triangles where none of the included baselines were pistoned, with a typical closure phase shown in Figure \ref{fig:badCP_nopiston}. 

\begin{figure}[ht!]
\centering\includegraphics[width=\linewidth]{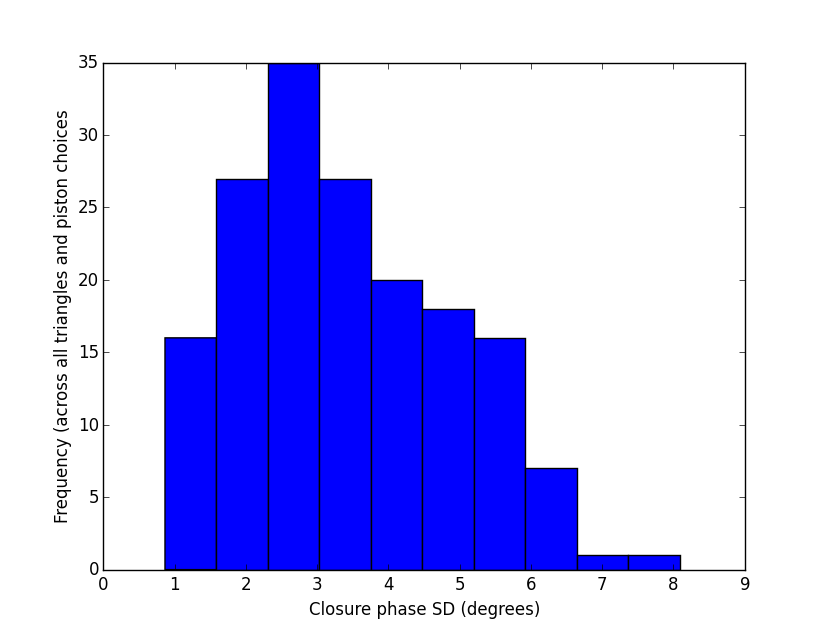}
\caption{The measured closure phase stability across all possible triangles, over a 6~$\mu m$ OPD piston ($8\pi$).}
\label{fig:CPstability}
\end{figure}

Measuring a changing phase for outputs where no phase piston was applied should not be possible, however the periodic nature of the closure phase instabilities found in Fig. \ref{fig:badCP} and \ref{fig:badCP_nopiston}, are indicative that the primary cause is cross-coupling between the waveguides. This can either arise from direct cross-coupling, where waveguides are brought too close to one another (for example at one of the many crossing points in the chip), or alternatively by interfering with the quasi-coherent stray light in the bulk arising from internal losses. The full characterization of closure phase stability for all possible closure-triangles on the chip is shown in Figure \ref{fig:CPstability}. For comparison, the closure phase stabilities measured of only the PRM in isolation, using free-space Fizeau interferometry of the output, was between $CP\sigma = 0.22^{\circ} - 1.0^{\circ}$ \cite{DflyBarn}.

While cross-coupling is a known, and on some level, expected phenomenon in photonic circuitry of such complexity, the important aspect is that the closure-phase instabilities were caused by a very small amount of cross-coupling ($1\%$ at maximum). For example, a cursory examination of Figure \ref{fig:waterfall} shows that the only modulations observed are in the baselines where the piston is applied, with the unpistoned baselines showing no visible change at first glance. It is not until the data is fitted to and the relative phase analyzed that the impact of cross-coupling becomes evident. For interferometric applications, this effect is therefore very important to consider when designing and fabricating photonic devices, as even minute quantities of cross-talk (which might be well below what is needed in other applications) can give rise to a decrease in performance. 

However, it is important to note that these are raw experimental stability measurements, not utilizing any of the calibration methods typically done on-sky, and over a piston error range much higher than in typical astronomical applications, especially behind AO systems. For integrated interferometric couplers such as this, their overall behaviour can be generalized into a matrix that includes not only all the waveguide combinations but also the incoherent and coherent cross-coupling terms (taking into account correlated error terms). This matrix is referred to as a Visibility to Pixel Matrix (V2PM), and can completely characterize the instrumental behaviour (which includes the cross-talk as well as the transmission, visibility, and phase) of the beam-combiner, and thus calibrate it out of the final measurements \cite{Tatulli_2006, Lacour_2008, Lapeyrere_2014}. The V2PM analysis method does not physically reduce the cross-coupling, but rather allows for it to be calibrated out of the final measurement. Nonetheless, the consistent closure phase stability of a few degrees across all 28 baselines simultaneously validates the use of photonic beam combiners interfaced with photonic PRMs, and is a significant step forwards. 


\section{Discussion}
\label{sec:discussion}
There are a number of methods to improve the overall throughput of the device that were not shown in this work. Firstly, the Fresnel reflections on the four air-glass interfaces in the assembly limit the maximum possible throughput to ($\sim84~\%$), and can easily be removed by adding commercially available anti-reflection coatings to the chip end-faces and the MLA. Secondly, while the MLA-PRM section of the device is already comparable in throughput to existing pupil-remapping interferometers deployed on telescopes \cite{FIRST1}, the injection of the light into the waveguides could potentially be improved using 3D printed lenslets on the chip surface \cite{HarrisMLA}. Another improvement can be made by using an alternative glass for the PRM (such as Schott AF45) that does not have the $\sim 20~\%$ absorption at NIR wavelengths. This glass was used to create similar devices with an internal throughput of $\sim 90\%$ as part of the GLINT interferometer \cite{GLINT1}.  

We can state with high confidence that the majority of the closure phase stability errors arise inside the beam-combining chip. This is because the MLA-PRM section was tested independently in the setup, with the beam-combination done in free-space, with a measured closure phase stability of $CP\sigma = \sim0.2^{\circ}$ \cite{DflyBarn}. The periodic nature of the error is consistent with that seen in devices where the light in the waveguides is interfering with stray unguided light in the bulk of the chip. While the PRM outputs are sidestepped, and a mask placed in the path to ensure none of the PRM stray light enters the beam-combining chip, the beam-combining chip itself has no such sidestep. Thus, light that is lost in the PRM-beam combiner interface freely travels through the entire chip to the output, and can interfere with the guided light. Further, any losses in the y-junction splitters or directional couplers also propagates towards the output waveguides and can similarly interfere. The easiest solution to this is to include either a sidestep of the input waveguides or an `\textit{around-the-corner}' $90^{\circ}$ bend design when making the beam-combiner. However, this would likely increase the overall chip footprint, making such a modification perhaps better suited to higher refractive index fabrication methods (such as Silicon Nitrite platforms). Nevertheless, this type of cross-coupling is common in photonic interferometers for astronomy, and robust analysis methods exist to calibrate correlated error terms. 

When considering possible on-sky performance of this device it is important to note that the overall range of the piston errors used in laboratory characterization is considerably larger than what is likely to be experienced on-sky, especially behind AO. Our applied piston values ranged uniformly from 0 to 2$\pi$ radians, corresponding to an RMS wavefront error of 1.8 radians -- extremely large compared to typical RMS wavefront error residuals of $\sim 80~nm$ or 0.3~radians for modern extreme-AO systems. The estimated closure phase stability of our device under an AO-corrected phase error regime is

\begin{equation}
CP\sigma_{on-sky} = \sqrt{3}\cdot \frac{\varepsilon_{AO}}{1.8}\cdot CP\sigma_{lab} 
\end{equation}

where $\sigma_{lab}$ is the experimentally measured closure phase stability and $\varepsilon_{AO}$ is the AO corrected phase error (in radians). The factor of $\sqrt3$ comes from the fact that in our lab, measurements only 1 baseline were affected by wavefront error, but on sky all 3 baselines in the triangle would suffer \cite{DflyBarn}. For our most stable triangle, this translated to a 
$CP\sigma_{on-sky} = 0.26^{\circ}$,
and for the average for all triangles being 
$CP\sigma_{on-sky} = 1.36^{\circ}$.
Using the relationship between closure phase precision and detectable contrast ratio given in \cite{CPtoCont}, this results
in a raw contrast-ratio sensitivity limit at $1\lambda/D$ ($1\sigma$ detection) of 
$6.5\times 10^{-4}$
and 
$34\times 10^{-4}$ respectively.

Lastly, while spectral dispersion was not used in these experiments, any on-sky implementation would likely cross-disperse the outputs to get multiple wavelength channels simultaneously. This technique not only provides additional scientific benefits, such as differential closure phase measurements at emission lines of interest, but also provides an additional parameter space for model fitting and handling of correlated errors.


\section{Conclusion}
We demonstrate the successful hybridization of two key technologies used in astrophotonic interferometry; a ULI fabricated 3D pupil remapping chip, and a lithographic pairwise ABCD beam-combiner. Further, we demonstrate a packaging solution to minimize on-telescope alignment by bonding the chips with the injection microlens array into a robust housing. This method of hybridization is a critical step for scaling photonic interferometers to many more inputs, particularly for making use of more of the pupil in the upcoming ELT era. This 8-input pairwise beam combiner is, to the authors' knowledge, the largest to-date. 

Under laboratory testing conditions, we measure the closure phase stability to be $CP\sigma = 0.9^{\circ}$ for the best case, over an $8\pi$ piston error range at the input. Other baseline combinations performed worse due to cross-coupling between the waveguides inside the beam-combining chip, with an average closure phase stability $CP\sigma = 4.7^{\circ}$. The total measured end-to-end throughput of the entire assembly was determined to be $\sim 26 \%$.









\begin{backmatter}
\bmsection{Acknowledgments}
This research was  supported by the Australian Research Council Centre of Excellence for Ultrahigh bandwidth Devices for Optical Systems (project number CE110001018). The work was performed in part at the OptoFab node of the Australian National Fabrication Facility utilizing Commonwealth as well as NSW state government funding. S. Gross acknowledges funding through a Macquarie University Research Fellowship (9201300682) and the Australian Research Council Discovery Program (DE160100714). N. Cvetojevic acknowledges funding through the European  Research Council (ERC) under the European Union’s Horizon 2020 research and innovation program (grant agreement CoG - 683029).
      
\bmsection{Disclosures}
The authors declare no conflicts of interest.

\bmsection{Data availability} Data underlying the results presented in this paper are not publicly available at this time but may be obtained from the authors upon reasonable request.

\end{backmatter}

\bibliography{references}

\bibliographyfullrefs{references}


\ifthenelse{\equal{\journalref}{aop}}{%
\section*{Author Biographies}
\begingroup
\setlength\intextsep{0pt}
\begin{minipage}[t][6.3cm][t]{1.0\textwidth} 
  \begin{wrapfigure}{L}{0.25\textwidth}
    \includegraphics[width=0.25\textwidth]{john_smith.eps}
  \end{wrapfigure}
  \noindent
  {\bfseries John Smith} received his BSc (Mathematics) in 2000 from The University of Maryland. His research interests include lasers and optics.
\end{minipage}
\begin{minipage}{1.0\textwidth}
  \begin{wrapfigure}{L}{0.25\textwidth}
    \includegraphics[width=0.25\textwidth]{alice_smith.eps}
  \end{wrapfigure}
  \noindent
  {\bfseries Alice Smith} also received her BSc (Mathematics) in 2000 from The University of Maryland. Her research interests also include lasers and optics.
\end{minipage}
\endgroup
}{}

\end{document}